\documentclass[aps,prb,twocolumn,floatfix]{revtex4}

\usepackage[svgnames]{xcolor}
\usepackage[colorlinks=true,urlcolor = teal, citecolor = blue, linkcolor= magenta, bookmarks=false]{hyperref}
\usepackage{amsfonts,amsmath,amssymb}
\usepackage{graphicx}
\usepackage{dsfont}
\usepackage{relsize}
\usepackage{color}
\usepackage{braket}
\usepackage{bm}

\DeclareMathOperator{\im}{Im}

\DeclareMathOperator{\Tr}{Tr}

\def\la{\label}

\def\be{\begin{equation}}
\def\ee{\end{equation}}

\newcommand{\bei}{\begin{itemize}}
\newcommand{\eei}{\end{itemize}}

\newcommand{\bee}{\begin{enumerate}}
\newcommand{\eee}{\end{enumerate}}

\def\a{\alpha}
\def\g{\gamma}

\def\e{\epsilon}

\def\s{\sigma}
\def\t{\tau}

\def\om{\omega}

\def\sfa{{\mathsf a}}

\def\cG{\mathcal G}

\def\cO{\mathcal O}
\def\cS{\mathcal S}
\def\cT{\mathcal T}

\def\cZ{\mathcal Z}

\definecolor{grey}{rgb}{0.4,0.4,0.5}
\definecolor{darkgreen}{rgb}{0,0.5,0}
\definecolor{darkred}{rgb}{0.6,0.0,0}
\definecolor{lightbrown}{rgb}{1,0.9,0.8}
\definecolor{brown}{rgb}{0.6,0.3,0.3}
\definecolor{darkblue}{rgb}{0,0,0.8}
\definecolor{darkmagenta}{rgb}{0.5,0,0.5}

\def\pa {\partial}

\def\su{\alg{su}}
\def\sus{{\su(2|2)}}
\def\us{{\alg{u}(2|2)}}

\def\su{{su}}
\def\sus{{\su(2|2)}}
\def\us{{{u}(2|2)}}

\def\OB{{\bm \phi}}
\def\OF{{\bm \psi}}
\def\OHu{{\bm \theta}}

\def\OA{{\bm a}}

\def\Oc{{\bm c}}
\def\Ocd{{\bm c}^\dagger}
\def\Oq{{\bm q}}

\def\OQ{{\bm q}}
\def\OQd{{\bm q}^\dagger}

\def\On{{\bm n}}

\def\OSi{{\bm \Sigma}}
\def\Os{{\bm s}}
\def\OS{{\bm S}}
\def\Oet{{\bm \eta}}

\def\OH{{\bm H}}

\def\vOS{\vec{{\bm S}}}
\def\vOs{\vec{{\bm s}}}
\def\vOSi{\vec{{\bm \Sigma}}}
\def\vOet{\vec{\bm \eta}}
\def\ssu{{\mathsmaller{\uparrow}}}
\def\ssd{{\mathsmaller{\downarrow}}}

\def\qe{{\circ}}
\def\qd{{\bullet}}

\def\inft{0^+}

\def\ii{\mathrm i}

\def\DV{\nabla}
\def\vol{\mathcal V}

\def\GT{\mathcal G}

\def\sou{\zeta}

\def\gT{\mathfrak g}
\def\wT{\Omega}
\def\Sg{\Sigma}

\begin{document}

\title{Fermi surface enlargement on the Kondo lattice}

\author{Eoin Quinn}
\email{eoin.quinn@u-psud.fr}
\affiliation{LPTMS, CNRS, Univ.~Paris-Sud, Universit\'e Paris-Saclay, 91405 Orsay, France}

\author{Onur Erten}
\email{onur.erten@asu.edu}
\affiliation{Department of Physics, Arizona State University, Tempe, AZ 85281, USA}

%\email{E.P.Quinn@uva.nl}
%\affiliation{Institute for Theoretical Physics, University of Amsterdam, Science Park 904, 1090 GL Amsterdam, The Netherlands}

\date{\today}

\begin{abstract}
The Kondo lattice model is a paradigmatic model for the description of local moment systems, a class of materials exhibiting a range of strongly correlated phenomena including heavy fermion formation, magnetism, quantum criticality  and unconventional superconductivity. Conventional theoretical approaches invoke fractionalization of the local moment spin through large-$N$ and slave particle methods. In this work we develop a new formalism, based instead on non-canonical degrees of freedom. We demonstrate that the graded Lie algebra $\su(2|2)$ provides a powerful means of organizing correlations on the Kondo lattice through a splitting of the electronic degree of freedom, in a manner which entwines the conduction electrons with the local moment spins. This offers a novel perspective on heavy fermion formation. Unlike slave-particle methods, non-canonical degrees of freedom generically allow for a violation of the Luttinger sum rule, and we interpret recent angle resolved photoemission experiments on Ce-115 systems in view of this. 
\end{abstract} 

\maketitle 

%\newpage

%\section*{Comments}
%\begin{itemize}
%\item We are presenting a new theory, we need to be sure to adequately describe the scope of existing theories.
%\end{itemize}

\section{Introduction} 
Metals with local moments provide a rich playground to study unconventional phases and quantum phase transitions. A range of interesting phenomena,  including unconventional superconductivity and non-Fermi liquid behavior, arise from competition between magnetism and the Kondo effect\cite{Doniach_1977,Stewart_RMP1984, Si_Science2010, Coleman_book, Wirth_NRM2016}. When magnetism wins and the local moments order, the electrons are free to form a canonical Fermi liquid. 
When the Kondo effect dominates, the local moment is quenched by the conduction electrons, giving rise to `heavy' electronic quasi-particles with effective masses as large as 1000$m_e$\cite{Andres_PRL1975}. This heavy fermion state is also characterized by an enlargement of the Fermi surface, observed in Hall conductivity\cite{Friedemann_PNAS2010, Custers_Nature2003}, magnetostriction\cite{Gegenwart_Science2007}, quantum oscillation\cite{Shishido_JPSJ2005} and angle resolved photoemission (ARPES) \cite{Fujimori_JoPCM2016, Si_arXiv2018, Chen_PRB2017, Chen_PRB2018, Chen_PRL2018} experiments. These two regimes are generally separated by critical behavior  associated with Kondo breakdown, which manifests itself as a non-Fermi liquid fan extending to finite temperatures\cite{Si_Nature2001, Coleman_JoPCM2001, Senthil_PRL2003, Pepin_PRL2007, Pepin_PRL2007_2, Paschen_Nature2004}.

A precise estimation of the enlargement of the Fermi volume $\mathcal{V}_{FS}$ requires a complete mapping of the Fermi surface, a challenging task  only very recently achieved with ARPES\cite{Chen_PRB2017, Chen_PRB2018, Chen_PRL2018}. The conclusions are remarkable: CeCoIn$_5$\cite{Chen_PRB2017}, CeIrIn$_5$\cite{Chen_PRB2018} and CeRhIn$_5$\cite{Chen_PRL2018} all show  enlargement which is significantly smaller than the anticipated $\mathcal{V}_{FS} \propto n_c+n_f$, where $n_c$ and $n_f$ are the conduction electron and local moment densities respectively.  For instance in CeCoIn$_5$\cite{Chen_PRB2017} the enhancement is only $\mathcal{V}_{FS} \propto n_c+0.2\pm0.05$. These observations suggest a violation of Luttinger's sum rule, a direct proportionality between electron density and Fermi surface volume which has been established for a canonical Fermi liquid \cite{Luttinger_PR1960, Luttinger_PR1960_2, Oshikawa_PRL2000}.

Although the Kondo impurity problem is exactly solvable\cite{Wilson_RMP1975, Andrei_PRL1980, Wiegmann_JoPSSP1981}, there is no exact solution for the Kondo lattice model. The standard analytic approaches such as large-$N$ employ fractionalization of the local moment spin\cite{Coleman_PRB1984, Senthil_PRL2003, Millis_JAP1987}. Within this approach there are two possibilities for the Fermi surface volume: (i) $\mathcal{V}_{FS} \propto n_c$  when there is no Kondo hybridization which occurs at high temperature, (ii) $\mathcal{V}_{FS} \propto n_c +n_f$ once the Kondo hybridization sets in. This enlargement of the Fermi surface is attributed to the local moment spin becoming  delocalized, thereby gaining  charge in relation to Luttinger's sum rule. Dynamical mean-field theory\cite{Georges_RMP1996, Si_JPSJ2014}, which is exact in infinite dimensions, goes beyond the large-$N$ mean-field description by introducing finite lifetime effects, but is in qualitative agreement with respect to Luttinger's sum rule.

It is worth highlighting that these systems are not the only cases where evidence for the violation of Luttinger's sum rule is observed. Another prominent example is the pseudogap regime of the cuprates, where quantum oscillation and Hall and thermal conductivity experiments indicate the existence of a Fermi surface whose volume drops to zero as half-filling is approached \cite{doiron2007quantum,Badoux_2016,PhysRevX.8.041010}. The analogy can be strengthened by drawing a parallel  between the non-Fermi liquid behavior appearing between the small and large Fermi surface regimes in local moment systems with that occurring between the Fermi liquid and pseudogap regimes in the cuprates \cite{Keimer_rev}.
Linking rearrangement of the Fermi surface and non-Fermi liquid behavior  offers a promising paradigm for characterising the phase diagram of strongly correlated electronic matter.

In this article we develop a novel theoretical framework for local moment systems. We demonstrate that the degrees of freedom of local moment systems can be reinterpreted through the non-canonical graded Lie algebra $\sus$, and exploit this to obtain a systematic description of strongly correlated behaviour. The resulting regime can be interpreted as a splitting of the electronic degree of freedom\cite{Quinn_PRB2018}, and exhibits a self-hybridization of the band structure inducing a heavy effective mass and enlargement of the Fermi surface.

Our formalism violates Luttinger's sum rule quite generally.   Central to Luttinger's theorem is the organization of the correlations of an interacting system around canonical fermion degrees of freedom via the Scwinger--Dyson equation, let us cast it as $\GT=\tfrac{1}{\GT_0^{-1}-\Sigma}$. Recently it has been established that correlations can instead be organized around non-canonical degree's of freedom via an exact representation of the Green's function as $\GT=\gT\wT=\tfrac{\wT}{\gT_0^{-1}-\Sigma}$, where $\wT$ encodes the correlations resulting from the non-canonical nature of the degree of freedom\cite{Shastry_PRL2011, Shastry_PRB2013}. In the purely electronic setting it was shown that this generically yields a violation of Luttinger's sum rule \cite{Quinn_PRB2018}. This can be regarded as formalising the operatorial approach put forward by Hubbard \cite{Hubbard_ProcRoySoc1964,Hubbard4}, as well as providing a framework for systematically going beyond it.

\section{Local moment systems}

We consider the  Kondo lattice Hamiltonian 
\be\la{KondoH}
\OH = \sum_{p, \s} (\varepsilon_p-\mu) \Ocd_{p\s} \Oc_{p\s} 
	+ J_K \sum_i \vOs_i\cdot \vOS_i,
\ee
an archetypal model to describe local moment physics in which itinerant electrons interact with local spin moments at each site of the lattice through a Kondo coupling. Here  $\vOs$ denotes the conduction electron spin 
\be\la{spin}
 \Os^z=\tfrac{1}{2}(\On_{\ssu}-\On_{\ssd}), ~~\Os^+ =\Ocd_{\ssu} \Oc_{\ssd},~~ \Os^- = \Ocd_{\ssd} \Oc_{\ssu},
\ee
and $\vOS$ denotes local moment spin.
We consider the general case of a spin-$S$ local moment, and so the Hilbert space at each site is $4(2S+1)$ dimensional. For example, for the case of a spin-$1/2$ local moment there are 8 states per site: $\ket{\ssd},~\Ocd_{\ssd}\ket{\ssd},~\Ocd_{\ssu}\ket{\ssd},~\Ocd_{\ssd} \Ocd_{\ssu}\ket{\ssd},~
\ket{\ssu},~\Ocd_{\ssd}\ket{\ssu},~\Ocd_{\ssu}\ket{\ssu},~\Ocd_{\ssd} \Ocd_{\ssu}\ket{\ssu}$.

In the absence of the Kondo coupling,  when $J_K=0$, the electrons and local moments 
are decoupled. For  $J_K\neq0$ however the interaction induces correlations in the system,  
and our objective is to identify those which allow for a  good effective description of the resulting behavior. Heuristically, we wish to identify the relevant degrees of freedom, and organize the correlations about these. In practice, a quantum degree of freedom is specified by the algebra it obeys, and this algebra provides the mathematical structure for organizing the correlations induced by the interacting Hamiltonian.

Let us outline two distinct ways of characterising the local degree of freedom. Firstly, the standard way is to regard the electrons and spin moments independently. Here the electrons are governed by the canonical anti-commutation relations $\{\Oc_{\s},\Ocd_{\s'}\}=\delta_{\s\s'}$, and the local spin moments are governed by the $\su(2)$ algebra $ [\OS^z,\OS^\pm]=\pm\OS^\pm$, $[\OS^+,\OS^-]=2\OS^z$. These provide reasonable degrees of freedom for a regime of behavior where the electrons form a Fermi liquid with a `small' Fermi surface, and the spins are free to order at low temperatures, as seen for example in CeRh$_2$Si$_2$ \cite{Pourret_JPSJ2017}. 

In this article we pursue a distinct description of the local degree of freedom. This builds upon recent work arguing that the graded Lie algebra $\su(2|2)$ is a valid degree of freedom for organizing correlations in the purely electronic setting \cite{Quinn_PRB2018}. 
The $\su(2|2)$ algebra admits a family of $4(2S+1)$-dimensional representations \cite{Beisert07,Arutyunov_2008}, which have a natural interpretation as combining a local spin moment with the electron. 
Let us consider  fermionic operators  written explicitly in terms of $\Oc$ and $\OS$ as follows
\be\la{defQ}
\begin{split}
&\OQd_{\ssd\qe} = \tfrac{1}{2}\Oc_{\ssu} + \tfrac{\lambda}{2S+1} \big( \tfrac{1}{2}\Oc_{\ssu} -\On_{\ssd}\Oc_{\ssu} + \Oc_{\ssd} \OS^- + \Oc_{\ssu} \OS^z\big),\\
&\OQd_{\ssu\qe} =\tfrac{1}{2}\Oc_{\ssd} +  \tfrac{\lambda}{2S+1} \big( \tfrac{1}{2}\Oc_{\ssd} - \On_{\ssu}\Oc_{\ssd} + \Oc_{\ssu} \OS^+ - \Oc_{\ssd} \OS^z\big),\\
&\OQd_{\ssd\qd} = \tfrac{1}{2}\Ocd_{\ssd} - \tfrac{\lambda}{2S+1} \big( \tfrac{1}{2}\Ocd_{\ssd}- \On_{\ssu}\Ocd_{\ssd} + \Ocd_{\ssu} \OS^- - \Ocd_{\ssd} \OS^z\big),\\
&\OQd_{\ssu\qd} =-\tfrac{1}{2}\Ocd_{\ssu} + \tfrac{\lambda}{2S+1} \big( \tfrac{1}{2}\Ocd_{\ssu} -\On_{\ssd}\Ocd_{\ssu} + \Ocd_{\ssd} \OS^+ + \Ocd_{\ssu} \OS^z\big).
\end{split}
\ee
These are related back to the canonical fermion operators through
\be\la{splitting}
\begin{split}
\Ocd_{\ssd} =\OQ_{\ssu\qe}+\OQd_{\ssd\qd},\quad
\Ocd_{\ssu} =\OQ_{\ssd\qe}-\OQd_{\ssu\qd},
\end{split}
\ee
and so we refer to this as a splitting of the electron, as in the electronic case. 

Let us examine the algebra they generate. 
Firstly, the anti-commutation relations of the $\OQ$ are
\be\la{FF}
\begin{split}
&  \{ \OQ_{\s\nu}, \OQd_{\s\nu}\}= \tfrac{1+\lambda^2}{4}+\tfrac{\lambda}{2S+1}(\nu \Oet^z - \s \OSi^z),\\
& \{\OQ_{\ssd\nu}, \OQd_{\ssu\nu}\}=\tfrac{\lambda}{2S+1}{ \OSi}^+, ~~~~~~~ 
	\{ \OQ_{\s\qe},\OQd_{\s\qd}\}= \tfrac{\lambda}{2S+1}{ \Oet}^+,\\
& \{ \OQ_{\ssu\nu}, \OQd_{\ssd\nu}\}= \tfrac{\lambda}{2S+1}{ \OSi}^-, ~~~~~~~
	\{\OQ_{\s\qd}, \OQd_{\s\qe}\}= \tfrac{\lambda}{2S+1}{ \Oet}^-,\\
&  \{ \OQ_{\s\nu}, \OQ_{\s'\nu'}\}=\{ \OQd_{\s\nu}, \OQd_{\s'\nu'}\}=\tfrac{1-\lambda^2}{4}\e_{\s' \s}	\e_{\nu \nu'} ,
\end{split}
\ee
which generate the total spin operators
\be\la{Spin}
\vOSi=\vOs+\vOS,
\ee
combining the electronic  and local moment spin, and the electronic charge operators
\be\la{charge}
 \Oet^z=\tfrac{1}{2}(\On_{\ssu}+\On_{\ssd}-1), ~~\Oet^+ =\Ocd_{\ssd} \Ocd_{\ssu},~~ \Oet^- = \Oc_{\ssu} \Oc_{\ssd}.
\ee
In evaluating these anti-commutators the Casimir identity $\vOS\cdot\vOS = S(S+1)$ is used.
The commutation relations between the $\OQ$ and $\OSi$ are
\be\la{FB1}
\begin{split}
\lbrack\OSi^z , \OQd_{\s\nu} \rbrack = \tfrac{\s}{2} \OQd_{\s\nu},\qquad 
&	[\OSi^z , \OQ_{\s\nu} ] = - \tfrac{\s}{2} \OQ_{\s\nu}, \\
[\OSi^+ , \OQd_{\ssd\nu} ] = - \OQd_{\ssu\nu},\qquad	
&	[\OSi^+ , \OQ_{\ssu\nu} ] =  \OQ_{\ssd\nu},\\
[\OSi^- , \OQd_{\ssu\nu} ] = - \OQd_{\ssd\nu},\qquad	
&	[\OSi^- , \OQ_{\ssd\nu} ] =  \OQ_{\ssu\nu},
\end{split}
\ee
and between the $\OQ$ and $\Oet$ are
\be\la{FB2}
\begin{split}
\lbrack\Oet^z , \OQd_{\s\nu} \rbrack = \tfrac{\nu}{2} \OQd_{\s\nu},\qquad 
&	[\Oet^z , \OQ_{\s\nu} ] = - \tfrac{\nu}{2} \OQ_{\s\nu}, \\
[\Oet^+ , \OQd_{\s\qe} ] = \OQd_{\s\qd},\qquad	
&	[\Oet^+ , \OQ_{\s\qd} ] = - \OQ_{\s\qe},\\
[\Oet^- , \OQd_{\s\qd} ] = \OQd_{\s\qe},\qquad	
&	[\Oet^- , \OQ_{\s\qe} ] = - \OQ_{\s\qd}.
\end{split}
\ee
The $\OSi$ and $\Oet$ mutually commute, and each obeys an $\su(2)$ algebra
\be\la{BB}
\begin{split}
[\OSi^z,\OSi^\pm]=\pm\OSi^\pm, \qquad &[\OSi^+,\OSi^-]=2\OSi^z,\\
[\Oet^z,\Oet^\pm]=\pm\Oet^\pm, \qquad &[\Oet^+,\Oet^-]=2\Oet^z.
\end{split}
\ee
In this way the $\OQ$ generate the $\sus$ algebra whose algebraic relations are Eqs.~\eqref{FF} and \eqref{FB1}-\eqref{BB}.
Furthermore, the algebra is  extended to $\us$ by incorporating the generator
\be\la{Hu}
\OHu  =\tfrac{S+1}{2}\lambda -\tfrac{\lambda}{2S+1}\Big(\vec{\OSi} \cdot \vec{\OSi}+\tfrac{1}{3}\vec{\Oet}\cdot \vec{\Oet}\Big), 
\ee
which obeys
\be\la{eq:Hu}
\begin{split}
\lbrack \OHu, \OQd_{\s\nu} \rbrack &=  \tfrac{1+\lambda^2}{4 } \OQd_{\s\nu} +\tfrac{1-\lambda^2}{4 } \e_{\s\s'}\e_{\nu\nu'} \OQ_{\s'\nu'},\\
\lbrack \OHu, \OQ_{\s\nu} \rbrack &= - \tfrac{1+\lambda^2}{4 } \OQ_{\s\nu} - \tfrac{1-\lambda^2}{4 } \e_{\s\s'}\e_{\nu\nu'} \OQd_{\s'\nu'},
\end{split}
\ee
and commutes with the $\OSi$ and $\Oet$.

The set of generators
\be\la{gens}
8\times\OQ,~~3\times\Os,~~3\times\Oet,~~ \OHu,
\ee 
thus offer a second way to characterise the local degree of freedom on the Kondo lattice. 
Our intention now is to regard these as composite operators, and to employ the algebra they obey to organize correlations so as to gain access to a strongly correlated regime of  behavior.  Their algebra is non-canonical, for example the anti-commutation relations of the $\OQ$ yield the generators of the spin and charge $\su(2)$ sub-algebras. This obstructs the use of canonical methods for evaluating two-point functions of the $\OQ$. 
The non-canonical terms however come with a prefactor $\tfrac{\lambda}{2S+1}$, and we will employ a formalism recently introduced by Shastry to organize the correlations they induce. A powerful consequence of the splitting of the electron, Eq.~\eqref{splitting}, is that once the two-point functions of the $\OQ$ are obtained then the electronic Green's function follows immediately through linear combinations.

To proceed, it is necessary to re-express the Kondo lattice model through the generators \eqref{gens}. The kinetic term becomes quadratic in $\OQ$, through the linearity of Eq.~\eqref{splitting}. The Kondo interaction  $\vOs \cdot \vOS$ can be re-expressed as quadratic in $\OSi$ and quartic in $\OQ$, as both $\Os$ and $\OS$ give terms quadratic in $\OQ$ through Eqs.~\eqref{spin},~\eqref{Spin}. It is however also possible to re-express the Kondo interaction in a simpler way. For this we rewrite Eq.~\eqref{Hu} using the operator identities $\vOs\cdot\vOs + \vOet\cdot\vOet  = \tfrac{3}{4}$ and $\vOS\cdot\vOS = S(S+1)$
to obtain
\be\la{Kint}
\vOs \cdot \vOS = \tfrac{1}{3}\vOet\cdot\vOet - \tfrac{2S+1}{2\lambda} \OHu - \tfrac{1-2S}{8}.
\ee
This convenient expression reflects the power of recasting the Kondo lattice model through $\su(2|2)$. It allows us to cleanly identify the role of the Kondo coupling in  splitting the electronic band, due to linear action of $\OHu$ on $\Oq$ from Eq.~\eqref{eq:Hu}.

%%%%%%%%%%%%%%%%%%%%%%%%%%%%%%%%%%%%%%%%%%%%%%
%%%%%%%%%%%%%%%%%%%%%%%%%%%%%%%%%%%%%%%%%%%%%%

\section{Organizing strong correlations}

We now exploit the $\su(2|2)$ algebra to gain access to a strongly correlated regime of behavior. Let us emphasise that we do not require the algebra $\su(2|2)$ to provide an explicit symmetry of the model in any way, instead we use it to organise correlations. Our ultimate objective is to compute the electronic Green's function
\be\la{thGFcan}
\begin{split}
{\GT^{\rm el}_{ij\s}}(\t)&= - \braket{\Oc_{i\s}(\t) \Ocd_{j\s}(0)}\\
	&=-\tfrac{1}{\cZ}\Tr \Big(e^{-\beta \OH}\cT\big[\Oc_{i\s}(\t) \Ocd_{j\s}(0)\big]\Big),
\end{split}
\ee
where $\cZ=\Tr e^{-\beta \OH}$, $\beta$ is inverse temperature, $\OA(\t)= e^{\t\OH}\OA e^{-\t\OH}$, and $\cT$ is the $\t$-ordering operator which is antisymmetric under interchange of fermionic operators.

This section closely mirrors Sec. III of Ref.~\cite{Quinn_PRB2018} where a corresponding analysis is made in the purely electronic setting. We adopt a simplifying notation, collecting the fermionic generators as
\begin{equation}
\OF_i^\a = \left(\begin{array}{cccccccc}
\OQd_{i\ssu\qe}&\OQ_{i\ssd\qd}& \OQd_{i\ssd\qe}&\OQ_{i\ssu\qd}&
 \OQ_{i\ssu\qe}&\OQd_{i\ssd\qd}&\OQ_{i\ssd\qe}&\OQd_{i\ssu\qd}
 \end{array}\right),
\end{equation}
with greek indices, and the bosonic generators as
\be
 \OB_i^a = \left(\begin{array}{cccccc}
 \OSi_i^z&\OSi_i^-&\OSi_i^+ &\Oet_i^z&\Oet_i^-&\Oet_i^+ 
\end{array}\right),
\ee 
with latin indices. The  $\us$ algebra is then compactly expressed as 
\be\la{u22alg}
\begin{split}
\{ \OF_i^\a,\OF_j^\beta\} & = \delta_{ij} \big(  f^{\a\beta}{}_I + f^{\a\beta}{}_a \OB_i^a\big),\\
 [ \OB_i^a,\OF_j^\beta]  &= \delta_{ij}f^{a \beta}{}_\g \OF_i^\g,\qquad
 [ \OB_i^a,\OB_j^b]  = \delta_{ij}f^{ab}{}_c \OB_i^c,\\
  [ \OHu_i,\OF_j^\a]  &= \delta_{ij}f^{\Theta \a}{}_\beta \OF_i^\beta,\qquad 
  [ \OHu_i,\OB_j^a]  =0,
\end{split}
\ee
where summation over repeated algebraic indices is implied. Explicit expression for the structure constants $f$ can be read from Eqs.~\eqref{FF} and \eqref{FB1}-\eqref{BB},  and given explicitly in Appendix~\ref{app:compact}.

The Kondo lattice Hamiltonian can then be re-expressed in terms of the split-electron degrees of freedom
\be\la{hamTTR}
\begin{split}
\OH &= - \sum_{\braket{i,j}} t_{ij,\a\beta} \OF_i^\a  \OF_{j}^\beta 
	+\sum_{i} V_{ab}  \OB_i^a \OB_i^b \\
	&\qquad+\sum_i V_\Theta \OHu_i -  \mu_{a} \sum_i \OB_i^a.
\end{split}
\ee
Here $\braket{i,j}$ denotes the summation is over pairs of sites, and the non-zero hopping parameters are $t_{ij,51}=t_{ij,61}=t_{ij,52}=t_{ij,62}=t_{ij,73}=-t_{ij,83}=-t_{ij,74}=t_{ij,84}=t_{ij}$ and their anti-symmetric pairs $t_{ij,\a\beta}=-t_{ij,\beta\a}$, where $t_{ij}=-\tfrac{1}{\vol}\sum_p e^{\ii p(i-j)}\varepsilon_p$ with $\vol$ the total number of lattice sites. The remaining non-zero parameters are  
$V_{44} = 2 V_{56} = 2 V_{65} = \tfrac{1}{3} J_K$, $V_\Theta= -\tfrac{2S+1}{2\lambda} J_K$ and $\mu_4 =  2 \mu$.

We set ourselves the intermediate objective of computing the
 matrix Green's function of the $\OQ$, that is
\be
\GT_{ij}{}^\a_\beta (\t,\t')=  -{\braket{\OF_{i}^\a(\t) \OF_{j\beta}(\t')}},
\ee
where $
\OF_{i\a} = \big(\OF_i^\a\big)^\dagger= \OF_i^\beta K_{\beta\a}$, which defines $K$ given explicitly in Appendix~\ref{app:compact}.
The electronic Green's function is  immediately  obtained from linear combinations of these
\be\la{GeGQ}
\begin{split}
\GT^{\rm el}_{ij\ssd}(\t) &= {\GT_{ij}}^1_1(\t)+{\GT_{ij}}^1_2(\t)+{\GT_{ij}}^2_1(\t)+{\GT_{ij}}^2_2(\t),\\
\GT^{\rm el}_{ij\ssu}(\t) &= {\GT_{ij}}^3_3(\t)-{\GT_{ij}}^3_4(\t)-{\GT_{ij}}^4_3(\t)+{\GT_{ij}}^4_4(\t),
\end{split}
\ee
via Eqs.~\eqref{splitting}.

The challenge in computing $\GT$ is the non-canonical nature of the algebraic relations Eq.~\eqref{u22alg}, which obstructs the use of Wick's theorem. To proceed we follow Shastry\cite{Shastry_PRL2011, Shastry_PRB2013} and employ the Schwinger formalism, introducing sources for the bosonic generators $\OB$ into the imaginary-time thermal expectation value as follows
\be\la{ev_source}
 \braket{ \cO(\t)} = \frac{\Tr \Big( e^{-\beta  \OH} \cT \big[e^{\int_0^\beta d\t' \cS(\t')} \cO(\t) \big]   \Big)}{\Tr \big( e^{-\beta H} \cT [e^{\int_0^\beta d\t' \cS(\t')}]   \big)},
\ee
with $\cS(\t)  = \sum_i \sou_{ia}(\t) \OB^a_i(\t)$. Then bosonic correlations can be traded for functional derivatives through 
\be
\begin{split}
\braket{\OB^a_i(\t)\cO(\t')}&= 
	 \Big(\braket{\OB^a_i(\t)}+\DV_i^a(\t)\Big)\braket{\cO(\t')},
\end{split}
\ee
where $\DV_i^a(\t) =  \frac{{\delta} }{ {\delta} \sou_{ia}(\t^+)}$, and $\t^+=\t+0^+$ incorporates an infinitesimal regulator which ensures a consistent ordering when $\t=\t'$.

The matrix Green's function obeys the equation of motion
\be
\begin{split}
\pa_{\t} \GT_{ij}{}^\a_\beta(\t,\t') =  -\delta(\t-\t')\braket{\{\OF_{i}^\a(\t), \OF_{j\beta}(\t)\}}&\\
	 + \braket{[\cS(\t), \OF^\a_i(\t)]\OF_{j\beta}(\t')}&\\
	 - \braket{ [\OH,\OF^\a_i(\t)]\OF_{j\beta}(\t') }&,
\end{split}
\ee
together with the anti-periodic boundary condition $\GT_{ij}{}^\a_\beta(\beta,\t')=-\GT_{ij}{}^\a_\beta(0,\t')$. Evaluating the algebraic relations, it takes the form 
\begin{widetext}
\be\la{GFeqn}
\begin{split}
\sum_k \Big[ 
\delta_{ik}\Big(-\delta^\a_\g \pa_{\t} -f^{a\a}{}_\g \sou_{ia}(\t) 
 - \mu_a f^{a\a}{}_\g + V_\Theta f^{\Theta\a}{}_\g 
 	-f^{a\a}{}_\delta V_{ab}f^{b\delta}{}_\g
 	+ 2f^{a\a}{}_\g V_{ab} \big(\braket{\OB_i^b(\t)}  +  \DV_i^b(\t) \big)\Big) ~~~~~~~~~~~~~~~ &\\
+ f^{\a\delta}{}_I t_{ik,\delta\g}
	+f^{\a\delta}{}_a t_{ik,\delta\g} \big( \braket{\OB_i^a(\t)} 
	 + \DV_i^a(\t) \big)\Big] \GT_{kj}{}^\g_\beta(\t,\t')&\\
= \delta(\t-\t')\delta_{ij}\big(f^{\a\g}{}_I+ f^{\a\g}{}_a\braket{\OB^a_i(\t)}\big) K_{\g\beta}&.
\end{split}
\ee

The canonical way to proceed here is to invert $\GT$ via the Schwinger--Dyson equation, but this is obstructed by the non-trivial expectation value on the right-hand side. 
Here we bypass this difficulty by adopting Shastry's trick of factorising $\GT$ in two
\be\la{Shansatz}
\GT_{ij}{}^\a_\beta(\t,\t') = \sum_l \int_0^\beta d\t''  \gT_{il}{}^\a_\g(\t,\t'') \wT_{lj}{}^\g_\beta(\t'',\t').
\ee
Distributing the functional derivative in Eq.~\eqref{GFeqn} across these factors, and bringing the terms with the functional derivative acting on $\wT$ to the right-hand side,  a simplification can be made by exploiting the arbitrariness in the definition of $\wT$ to set
\be\la{eqn0}
\begin{split}
\wT_{ij}{}^\a_\beta(\t,\t') =& \delta(\t-\t')\delta_{ij} \big(f^{\a\g}{}_I +f^{\a\g}{}_a \braket{\OB^a_i(\t)} \big)K_{\g\beta}
	-\sum_{k,l}\int_0^\beta d\t'' \Big(f^{\a\e}{}_a t_{il,\e\delta} +2\delta_{il} f^{b\a}{}_\delta V_{ba} \Big) 
			\gT_{lk}{}^\delta_\g(\t,\t'') \DV_i^a(\t)\wT_{kj}{}^{\g}_\beta(\t'',\t').
\end{split}
\ee
The equation of motion then reduces to 
\be\la{gFeqn}
\begin{split}
\sum_k \Big[ 
\delta_{ik}\Big(-\delta^\a_\g \pa_{\t} -f^{a\a}{}_\g \sou_{ia}(\t) 
 - \mu_a f^{a\a}{}_\g + V_\Theta f^{\Theta\a}{}_\g 
	-f^{a\a}{}_\delta V_{ab}f^{b\delta}{}_\g
 	+ 2  f^{a\a}{}_\g V_{ab} \big(\braket{\OB_l^b(\t)} +  \DV_l^b(\t) \big)\Big) ~~~~~~~~~~~~~~~ &\\
+ f^{\a\delta}{}_I t_{ik,\delta\g}
	+f^{\a\delta}{}_a t_{ik,\delta\g} \big( \braket{\OB_i^a(\t)} 
	 + \DV_i^a(\t) \big)\Big] \gT_{kj}{}^\g_\beta(\t,\t')
	 = \delta(\t-\t')\delta_{ij}&.
\end{split}
\ee
We have thus converted Eq.~\eqref{GFeqn} with one unknown $\GT$ into two equations Eqs.~\eqref{eqn0},~\eqref{gFeqn} with two unknowns $\wT$, $\gT$. The advantage is that Eq.~\eqref{eqn0} is a closed functional equation for $\wT$, while Eq.~\eqref{gFeqn} has the form of a canonical equation of motion, and thus can be inverted through  the Scwhinger--Dyson equation in the standard way as follows
\be\la{eqn3}
\begin{split}
\gT^{-1}_{ij}{}^\a_\beta(\t,\t') &= \gT_{0,ij}^{-1}{}^\a_\beta(\t,\t') - \Sg_{ij}{}^\a_\beta(\t,\t'),
\end{split}
\ee
where $\gT_0$ is given exactly through
\be\la{eqn4}
 \Big[ 
\delta_{ik}\big(-\delta^\a_\g \pa_{\t}   -f^{a\a}{}_\g \sou_{ia}(\t)
 - \mu_a f^{a\a}{}_\g + V_\Theta f^{\Theta\a}{}_\g\big) + f^{\a\delta}{}_I t_{ik,\delta\g}
 		\Big] \gT_{0,kj}{}^\g_\beta(\t,\t')\\
	= \delta(\t-\t')\delta_{ij}\delta^\a_\beta,
\ee
and $\Sg$ obeys the closed functional equation
\be\la{eqn5}
\begin{split}
\Sg_{ij}{}^\a_\beta(\t,\t')  =& 
	\delta(\t-\t')\delta_{ij} f^{a\a}{}_\g V_{ab}f^{b\g}{}_\beta 
	- \delta(\t-\t')\Big(
	f^{\a\g}{}_a t_{ij,\g\beta}   +
	2\delta_{ij} f^{b\a}{}_\beta V_{ba} 
	\Big)\braket{\OB_i^a(\t)}\\
&- \delta(\t-\t')\delta_{ij}\sum_l\Big( f^{\a\e}{}_a  t_{il,\e\delta} 
 	+2\delta_{il}f^{b\a}{}_\delta V_{ba} \Big)\gT_{li}{}^\delta_\g(\t,\t^+)f^{a\g}{}_\beta \\
&-\sum_{k,l}\int_0^\beta d\t'' \Big(f^{\a\e}{}_a t_{il,\e\delta} +2\delta_{il} f^{b\a}{}_\delta V_{ba} \Big) 
			\gT_{lk}{}^\delta_\g(\t,\t'') \DV_i^a(\t)\Sg_{kj}{}^{\g}_\beta(\t'',\t').
\end{split}
\ee
\end{widetext}

In this way, we obtain an exact representation of $\GT$ through Eqs.~\eqref{Shansatz}-\eqref{eqn0},~\eqref{eqn3}-\eqref{eqn5}, via an exact rewriting of the equation of motion for $\GT$. While at first sight these expressions may appear complicated, conceptually they are quite simple. Schematically the Green's function of the $\OQ$ is cast in the form $\GT\sim \gT \wT\sim \tfrac{\wT}{\gT_0^{-1}-\Sg}$, where $\gT_0^{-1}$ is known exactly and both $\wT$ and $\Sg$ obey exact closed functional equations. The appearance of a non-trivial numerator here is intuitively understood as capturing the correlations resulting from the non-canonical nature of the degree of freedom.

In general we cannot solve these equations exactly, i.e. we cannot gain complete control of all correlations in the system. Instead we use them to organize the correlations: $\wT$ and $\Sg$ can be computed through a perturbative expansion in $\tfrac{\lambda}{2S+1}$ and $J_K$, under the principle  that the leading contributions capture the crucial correlations governing the behavior in the regime governed by these non-canonical degrees of freedom. In the following section we focus on the simplest non-trivial approximation, which is to suppress the terms containing functional derivatives in Eqs.~\eqref{eqn0},~\eqref{eqn5}.
This is the static approximation, the analogue of Hartree-Fock for a canonical degree of freedom, where both $\wT$ and $\Sg$ are frequency independent.

We conclude by highlighting a subtlety arising in the local moment setting which is absent in the purely electronic case, i.e.~for $S=0$. This concerns computing terms of the form $\braket{\OB}$ and $\DV\braket{\OB}$. In the electronic case the $\OB$ are quadratic in $\OF$, and so $\braket{\OB}$ is directly obtained from $\GT$. For $S\neq0$ however, it is not quite this simple. The spin generators are $\vOSi=\vOs+\vOS$, and while $\vOs$ is quadratic in $\OQ$, it is necessary to understand how to handle the contributions of the form $\braket{\OS}$ and $\DV\braket{\OS}$. In the following we focus on the normal state within an approximation for which this subtlety does not affect the analysis.

%%%%%%%%%%% FIGURE %%%%%%%%% FIGURE %%%%%%%%%%%%%%FIGURE %%%%%%%%% FIGURE %%%%%%%%%%%%%%%%
\begin{figure*}[t]
\centering
\includegraphics[width=2.05\columnwidth]{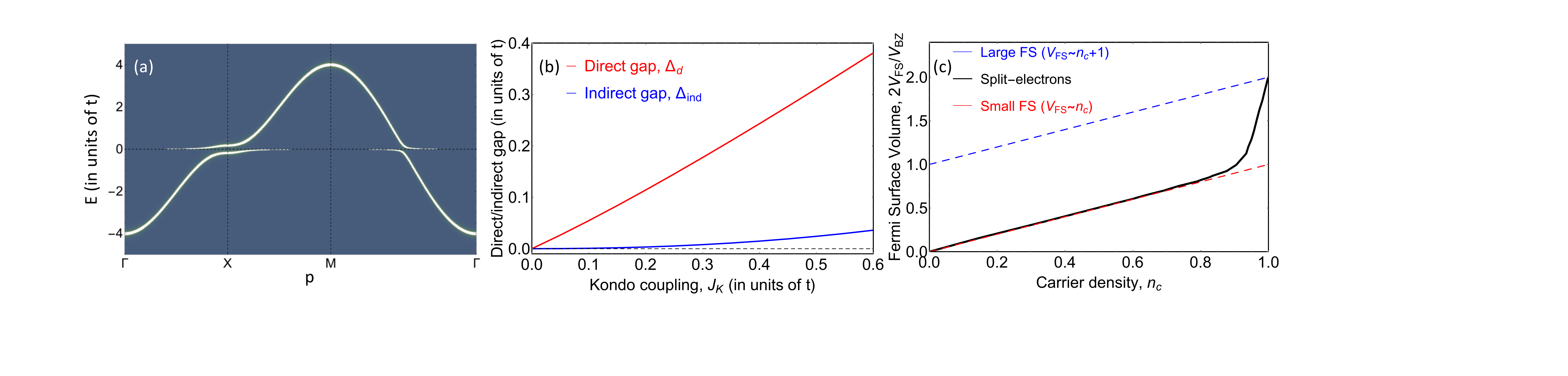}
\caption{\label{plot_SF}
%(Color online) 
Results from the static approximation within the split-electron formalism for a square lattice and  $S=1/2$:  (a) intensity plot of the electronic spectral function for $J/t=0.3$ and $\mu=0$ (with Lorentzian braodening) showing the formation of heavy bands around half-filling. (b) The direct and indirect gaps as a function of $J_K$, which are  related by $\Delta_{ind}=\Delta_d^2/W$ as in the large-$N$ mean-field approximation\cite{Coleman_book}. (c) Violation of Luttinger's sum rule. In contrast to standard theories where the Fermi surface volume is either $\mathcal{V}_{FS}\propto n_c$ or  $\mathcal{V}_{FS}\propto n_c+n_f$, we find $n_c \le 2\tfrac{\mathcal{V}_{FS}}{\mathcal{V}_{BZ}}\le n_c+1$, where $\mathcal{V}_{BZ}$ is the volume of the Brillouin zone.}
\end{figure*}
%%%%%%%%%%% FIGURE %%%%%%%%% FIGURE %%%%%%%%%%%%%%FIGURE %%%%%%%%% FIGURE %%%%%%%%%%%%%%%% 

\section{Static approximation}

We proceed to study the static approximation to the Green's function resulting from an organization of the correlations around the split-electron $\su(2|2)$ degrees of freedom.
This amounts to neglecting the functional derivative terms in Eqs.~\eqref{eqn0},~\eqref{eqn5}, which are suppressed in $\tfrac{\lambda}{2S+1}$ and $J_K$. We focus on the normal state, and so the only possible non-zero $\braket{\OB_i(\t)}$ is 
$\braket{\Oet^z_i(\t)}=\varphi^\a_\beta \GT_{ii}{}^\beta_\a(\t,\t^+)$, with $\varphi$  given explicitly in Appendix~\ref{app:compact}.

We thus set the sources to zero and switch to Fourier space according to 
\be\la{FT}
\cG_{p\s}(i\om_n) = \frac{1}{\vol}\sum_{i,j}\int_0^{\beta}  d\t  e^{\ii \om_n\t-\ii p(i-j)} \cG_{ij\s}(\t),
\ee
with Matsubara frequencies $\om_n=(2n+1)\frac{\pi}{\beta}$, $n\in\mathbb Z$, and $\vol$ is the total number of lattice sites. Then Eqs.~\eqref{Shansatz}-\eqref{eqn0},~\eqref{eqn3}-\eqref{eqn5} take the closed form
\be\la{eq:static}
\begin{split}
&\GT_p{}^\a_\beta(\ii\om_n) = \gT_{p}{}^\a_\g(\ii\om_n) \wT_{p}{}^\g_\beta,\\
&\wT_{p}{}^\a_\beta = \big(f^{\a\g}{}_I +f^{\a\g}{}_a \braket{\OB^a} \big)K_{\g\beta},\\
&\gT^{-1}_{p}{}^\a_\beta(\ii\om_n) = \gT_{0,p}^{-1}{}^\a_\beta(\ii\om_n) - \Sg_{p}{}^\a_\beta,\\
&\gT^{-1}_{0,p}{}^\a_\beta(\ii\om_n) = 
\ii\om_n\delta^\a_\beta  - \mu_a f^{a\a}{}_\beta 
		+ V_\Theta f^{\Theta\a}{}_\beta + f^{\a\delta}{}_I t_{p,\delta\beta},\\
&\Sg_{p}{}^\a_\beta  = f^{a\a}{}_\g V_{ab}f^{b\g}{}_\beta - \big(
	f^{\a\g}{}_a t_{p,\g\beta}   +
	2f^{b\a}{}_\beta V_{ba} 
	\big)\braket{\OB^a}\\
&~~~~ - \frac{1}{\vol}\sum_{q}  \big( f^{\a\e}{}_a  t_{q,\e\delta} 
 	+2f^{b\a}{}_\delta V_{ba} \big)\bar{\gT}_{q}{}^\delta_\g f^{a\g}{}_\beta,
\end{split}
\ee
where here the non-trivial $\braket{\OB^a}$ is given by $\braket{\Oet^z}=\frac{1}{\beta\vol}\sum_{q,m}e^{\ii \om_m \inft} \varphi^\a_\beta\GT_{q}{}^\beta_\a(\ii\om_m)$ and $\bar{\gT}_{q}{}^\a_\beta= \frac{1}{\beta}\sum_{m} e^{\ii \om_m \inft}\gT_{q}{}^\a_\beta(\ii \om_m)$. The corresponding approximate electronic Green's function follows through Eq.~\eqref{GeGQ}.

To illustrate the formalism we consider the Kondo lattice model on a two-dimensional square lattice with nearest-neighbour hopping. We solve Eqs.~\eqref{eq:static} self-consistently, and focus on $J_K=0.3$, $S=1/2$ and zero temperature. Our results are independent of the value of $\lambda$ chosen. In Fig. 1(a) we plot the electronic spectral function $A^{\rm el}_{p\s} = -\tfrac{1}{\pi} \im \GT^{\rm el}_{p\s}(\om+\ii\inft)$, which reveals the formation of heavy bands with large effective masses in the vicinity of half-filling. Unlike large-$N$ theories, the hybridization does not follow the chemical potential as one moves away from half-filling, though this may be a limitation of the static approximation. Our band structure also does not display any noteworthy temperature dependence. Figure 1(b) displays both the direct $\Delta_{\rm d}$ and the indirect $\Delta_{\rm ind}$ gaps as a function of $J_K$ for $\mu=0$. Similar to large-$N$ calculations\cite{Coleman_book}, we find the two gaps are related as $\Delta_{\rm ind} = \Delta_{\rm d}^2/W$, where $W$ is the bandwidth. Generically we find that the enlargement of the Fermi surface violates Luttinger's sum rule, and this is illustrated in Fig. 1(c). 
While for low electron density $n_c$ the Fermi surface volume closely obeys $V_{FS}\propto n_c$, as half-filling is approached the volume grows rapidly to $V_{FS}\propto n_c+1$.

\section{Summary and Discussion}

In this article we have developed a novel framework capable of characterising a strongly correlated regime of behavior on the Kondo lattice. We have shown how the local degree of freedom can be recast through the graded Lie algebra $\sus$, which can be interpreted as a splitting of the electron $\Oc\rightarrow\Oq+\Oq$. To handle the non-canonical nature of the algebra we have utilised Shastry's Green's function factorization technique, which leads to an exact representation of the $\Oq$ Green's function, with correlations encoded through two functional Eqs.~\eqref{eqn0},~\eqref{eqn5}. The electronic Green's function follows immediately through Eq.~\eqref{GeGQ}.

To examine the behavior governed by the split electrons we have focused on the `static' approximation. This is a first order approximation, the analogue of Hartree--Fock  for a canonical degree of freedom, in which the quasi-particles are sharply defined as shown in Fig. 1(a). 
As a function of parameters, it is possible to have a Kondo insulator at half filling or a heavy Fermi liquid with large Fermi surface and heavy quasi-particles away from half-filling.  We thus see that this captures the basic phenomenology of heavy fermions.

In contrast with prominent theories of heavy fermion formation, our analysis does not invoke a `delocalization' of the local moment spin. We find this an attractive aspect of our formalism, as the effective Kondo lattice setting has the charge of the local moment frozen out to begin with. Instead the moment's spin is entwined with the conduction electrons into the $\Oq_{\s\nu}$ as in Eq.~\eqref{defQ}. The Kondo splitting of the electronic band arises from a hybridization between the two flavors $\Oq_{\s\qe}$ and   $\Oq_{\s\qd}$. The enlargement of the Fermi surface emerges naturally, and can be attributed to violation of  Luttinger's sum rule due to the non-canonical nature of the degrees of freedom.

Indeed, violation of the Luttinger sum rule is another attractive feature of our formalism, unambiguously distinguishing it from existing theoretical approaches. It accounts for recent ARPES studies which find that the enlargement of the Fermi surface in CeCoIn$_5$\cite{Chen_PRB2017}, CeIrIn$_5$\cite{Chen_PRB2018} and CeRhIn$_5$\cite{Chen_PRL2018} is significantly smaller than the volume $\mathcal{V}_{FS}\propto n_c+n_f$ corresponding to delocalized spin moments. Within the large-$N$ framework, a possible explanation would be that some of the $f$-electrons  remain localized in a spin liquid. There is however no direct evidence for such behavior in these compounds. 
For instance, a putative $U(1)$ spin liquid would lead to a spinon continuum in neutron scattering experiments, and this has not been observed. In contrast, the split-electron degrees of freedom form a sharp Fermi surface and therefore recovers Fermi liquid phenomenology including $\rho \sim T^2$ resistivity at low temperatures.

There are many directions for future research. Of particular importance is going beyond the static approximation considered here. For the single impurity case, we do not expect to capture Kondo resonance formation within the static approximation, in line with the conventional perspective \cite{Hewson_book}. This motivates the development of improved approximative schemes along the lines of $T$-matrix or RPA methods. Indeed, it is  remarkable that the static approximation captures the hybridization gap. Recent ARPES experiments\cite{Kummer_PRX2015, Chen_PRB2017} show that the temperature at which the hybridization gap starts to open can be much higher than the Kondo coherence temperature, and we anticipate that improved approximations can recover the Kondo resonance and shed light on this dichotomy.

Another direction is to address magnetism. Within the large-$N$ framework this is a significant challenge, and attempts in this direction have been to extend the theory to supersymmetic versions\cite{Pepin_1996, Coleman_NucPhysB2000, Coleman_PRB2000, Coleman_PhysicaB2000, Ramirez_PRB2016}. On the other hand, although there are subtleties to be addressed within our formalism regarding magnetism, we no not expect an inherent bottleneck.  It would be interesting to examine magnetism in underscreened Kondo model, where $S>1/2$, for instance in the context of Uranium based ferromagnets\cite{Schoenes_PRB1984, Bukowski_JAC2005, Perkins_EPL2007}.

We conclude with a general comment, mirroring a similar analysis in the purely electronic setting \cite{Quinn_PRB2018}. We have identified two distinct ways to characterise the local degree of freedom on the Kondo lattice, either in the traditional way through the canonical fermion and local spin algebras, or through the $\sus$ algebra as developed here. Neither provides an exact solution of the model away from $J_K=0$. Instead they offer two distinct quasi-particle frameworks for organizing the correlations induced by interactions. It would be interesting to explore to what extent the competition between these two descriptions is responsible for the non-Fermi liquid behavior associated with Kondo destruction.

\section{Acknowledgements}
We thank Piers Coleman and Filip Ronning for fruitful discussions. OE is supported by  ASU startup grant. This work is funded in part by a
QuantEmX grant from ICAM and the Gordon and Betty Moore Foundation through Grant GBMF5305 to Eoin Quinn.

\appendix

\section{Compact notations}\la{app:compact}

The structure constants for the representation of the $\us$ algebra in Eq.~\eqref{u22alg} are conveniently expressed through tensor products of Pauli matrices
$
\s_0=\begin{pmatrix} 
1 & 0 \\
0 & 1 
\end{pmatrix}$,
 $\s_1=\begin{pmatrix} 
0 & 1 \\
1 & 0 
\end{pmatrix}$,
$\s_2=\begin{pmatrix} 
0 & -\ii \\
\ii & 0 
\end{pmatrix}$,
$\s_3=\begin{pmatrix} 
1 & 0 \\
0 & -1 
\end{pmatrix}$.
Firstly, $f^{\a\beta}{}_I$ and  $f^{\Theta\a}{}_\beta$ depend on $\lambda$ through $\sfa_\pm=\tfrac{1\pm \lambda^2}{4}$ as follows
\be
\begin{split}
f^{\a\beta}{}_I&=\sfa_+ \s_1\otimes\s_0\otimes\s_0+\sfa_- \s_1\otimes\s_3\otimes\s_1,\\
f^{\Theta\a}{}_\beta &= \sfa_+ \s_3\otimes\s_0\otimes\s_3-\ii \sfa_- \s_3\otimes\s_3\otimes\s_2.
\end{split}
\ee
The structure constants $f^{\a\beta}{}_a$ are proportional to $\tfrac{\lambda}{2S+1}$ as follows 
 \be
\begin{split}
f^{\a\beta}{}_1&=-\tfrac{\lambda}{2S+1} \s_1\otimes\s_3\otimes\s_3, \\
f^{\a\beta}{}_2&=\tfrac{\lambda}{2S+1} \tfrac{\s_1\otimes\s_1+\s_2\otimes\s_2}{2}\otimes\s_0, \\
f^{\a\beta}{}_3&=\tfrac{\lambda}{2S+1} \tfrac{\s_1\otimes\s_1-\s_2\otimes\s_2}{2}\otimes\s_0, \\
f^{\a\beta}{}_4&=-\tfrac{\lambda}{2S+1} \s_1\otimes\s_0\otimes\s_3, \\
f^{\a\beta}{}_5&=\tfrac{\lambda}{2S+1} \tfrac{\s_0+\s_3}{2}\otimes\s_1\otimes\s_1, \\
f^{\a\beta}{}_6&=\tfrac{\lambda}{2S+1} \tfrac{\s_0-\s_3}{2}\otimes\s_1\otimes\s_1. 
\end{split}
\ee
 The structure constants $f^{a\a}{}_\beta$ are independent of $\lambda$ as follows 
 \be
\begin{split}
f^{1\a}{}_\beta & = \tfrac{1}{2}\s_3\otimes\s_3\otimes\s_0, \\
f^{2\a}{}_\beta & = - \tfrac{\s_3\otimes\s_1+\ii \s_0\otimes\s_2}{2}\otimes\s_3, \\
f^{3\a}{}_\beta & = - \tfrac{\s_3\otimes\s_1-\ii \s_0\otimes\s_2}{2}\otimes\s_3, \\
f^{4\a}{}_\beta & = - \tfrac{1}{2}\s_3\otimes\s_0\otimes\s_0, \\
f^{5\a}{}_\beta & = - \tfrac{\s_2+\ii \s_1}{2}\otimes\s_1\otimes\s_2,\\
f^{6\a}{}_\beta & = - \tfrac{\s_2-\ii \s_1}{2}\otimes\s_1\otimes\s_2. \\
 \end{split}
\ee
Also $K^\a_\beta = \s_1\otimes\s_0\otimes\s_0$ and 
$\varphi^\a_\beta= \tfrac{\s_3\otimes\s_3-\s_0\otimes\s_0}{4}\otimes\s_1 +	
	\tfrac{\s_3\otimes\s_0-\s_0\otimes\s_3}{4}\otimes\s_0$.

\bibliographystyle{apsrev4-1}
\bibliography{KLbib}

\end{document}